\documentclass[
superscriptaddress,
 amsmath,amssymb,
 pre,twocolumn
]{revtex4-1}

\usepackage{graphics}
\usepackage{graphicx}
\usepackage{dcolumn}
\usepackage{bm}

\graphicspath{{./}}

\begin{document}

\preprint{APS/123-QED}

\title{Effects of update rules on networked N-player trust game dynamics}

\author{Manuel Chica}
\email{manuel.chicaserrano@newcastle.edu.au}
\author{Raymond Chiong}%
 \email{raymond.chiong@newcastle.edu.au}
\affiliation{%
 School of Electrical Engineering and Computing, The University of Newcastle, 
 Callaghan, NSW 2308, Australia
}%

\author{Jos\'{e} J. Ramasco}%
 \email{jramasco@ifisc.uib-csic.es}
\affiliation{
 Instituto de F\'{i}sica Interdisciplinar y Sistemas Complejos IFISC (CSIC-UIB), \\
 07122 Palma de Mallorca, Spain
}

\author{Hussein Abbass}
 \email{h.abbass@adfa.edu.au}
\affiliation{%
 School of Engineering and Information Technology, University of New South Wales,
 Canberra, ACT 2610, Australia 
}%


\date{\today}

\begin{abstract}

We investigate the effects of update rules on the dynamics of an evolutionary game-theoretic model  --  the \textit{N}-player evolutionary trust game -- consisting of three types of players: investors, trustworthy trustees, and untrustworthy trustees. Interactions between players are limited to local neighborhoods determined by predefined spatial or social network topologies. We compare evolutionary update rules based on the payoffs obtained by their neighbors. Specifically, we investigate the dynamics generated when players use a deterministic strategic rule (i.e., unconditional imitation with and without using a noise process induced by a voter model), a stochastic pairwise payoff-based strategy (i.e., proportional imitation), and stochastic local Moran processes. We explore the system dynamics under these update rules based on different social networks and different levels of game difficulty. We observe that there are significant differences on the promoted trust and global net wealth depending on the update rule. If the game is harder, rules based on unconditional imitation achieve the highest global net wealth in the population. Besides a global perspective, we also study the spatial and temporal dynamics induced by the rules and we find important spatio-temporal correlations in the system for all rules. Indeed, the update rules lead to the formation of  fractal structures on a lattice and, when the rules are stochastic, also the emergence of low frequencies in the output signal of the system (i.e., long-term memory).



\end{abstract}

\maketitle


\section{\label{sec:introduction}Introduction}


Evolutionary game theory is a mathematical framework for investigating the dynamics of strategies in populations ranging from two players to structured societies~\cite{Nowak06}. Trust between players plays a major role in the evolution of games in a social context, and has deep implications for the collective action of social and human systems~\cite{Reina06, Grodzinsky15, Petraki14}. As players rely on trust to handle complex problems and make decisions, relationships are formed. These relationships, in turn, give rise to opportunities and allow trust to spread~\cite{Abbass16}. A recent study found that players are more trustworthiness when having uncalculating cooperation~\cite{Jordan16a}. 

In terms of models, the most well-known version of trust games considers interactions between two types of players: \textit{investors} (or \textit{trusters}) and \textit{trustees}~\cite{Berg95, Cox04, King05,Tarnita15}. In these games, the investor must first decide whether to trust the trustee. If the decision is positive, the trustee must then decide whether to be trustworthy or not. 
Although two-player games are the most common configuration for trust games in the literature, they have their limitations. For instance, we cannot generalize some of the insights found in pairwise interactions to games with multiple players having more than two strategies~\cite{Gokhale10}. 

Abbass et al.~\cite{Abbass16} therefore proposed an evolutionary \textit{N}-player trust game to generalize the concept of trust to a well-mixed population of individuals who play a trust game concurrently. 
Chica et al.~\cite{Chica17IEEETEC} subsequently showed that promoting trust in this \textit{N}-player trust game depends on the social network structure. In their model, a structured population of players have three possible strategies to choose from: to be an investor ($I$), a trustworthy trustee ($T$), or an untrustworthy trustee ($U$). Players interact with their direct neighbors on the social network, and obtain their payoff values repeatedly through a number of time steps. 
They evolve their strategies according to the obtained payoffs and update their strategies in a non-deterministic way (i.e., using a proportional imitation update rule). 

In this paper, we extend on the work of Chica et al.~\cite{Chica17IEEETEC} by studying the implications in terms of spatio-temporal correlations of different update rules in the same networked multi-player trust game. In addition to a proportional imitation update rule, we also consider unconditional imitation, a local Moran process, and a hybrid approach using unconditional imitation and a voter model in a stochastic function.
In three-strategy games, systems can exhibit repetitive succession of oscillatory and stationary states~\cite{Szolnoki09}. This is because when having three strategies, the outcome can be considerably more complicated and the game can end in ever-increasing oscillations or with the elimination of some of the existing strategies~\cite{Nowak06}. It is thus important for us to examine the effects of different update rules and how the system dynamics would change in this context.  

In our simulations, we consider different social network topologies (i.e., a simple regular lattice~\cite{Nowak92} and scale-free (SF) networks~\cite{Albert02,Barabasi99}). We also explore the effects of the update rules for different values of the temptation to defect ratio  $r_{UT}$ in the trust game. This ratio $r_{UT}$ defines the temptation to defect investors' trust by the trustees and measures the level of difficulty of the game. We evaluate the final steady state (i.e., surviving strategies) and global net wealth (i.e., total payoffs) of the population to observe which update rules and under which conditions players can better promote trust and obtain the highest net wealth of the population.

The effects of the different update rules are then examined to better understand how trustworthy and untrustworthy behaviors emerge and what is their corresponding global net wealth like. The dynamics of the model are first studied from the spatial perspective. It is known that spatial games of evolution can generate dynamic fractals, gliders and kaleidoscopes~\cite{Nowak92,Nowak06}, and so we analyze the spatial formations as well as those spatial correlations between the players' behaviors and strategies. Finally, we show the temporal dynamics of the model by analyzing the signal of the system and investigate if the majority of the rules generate those correlations. We also explore if more elaborated correlations with both spatial and temporal types occur in the model's output.

\section{\label{sec:model}MODEL}   

\subsection{\label{sec:model_desc} Model description}

The multi-player trust game consists of a finite set of agents occupying the nodes of a social network, with edges denoting interactions between them (both for accumulating payoffs and strategy updating~\cite{Nowak10}). Each player $i$ chooses a strategy $s$ from three possibilities at every time step ($s(i)$)~\cite{Chica17IEEETEC}: being an investor (strategy $I$), being a trustworthy trustee (strategy $T$), and being an untrustworthy trustee (strategy $U$). During the evolutionary game process, all players interact with their directly connected neighbors at the same time in a single game (group interaction~\cite{Chiong12TEC}). 

\begin{figure}
 \centering
\includegraphics[width=0.5\textwidth]{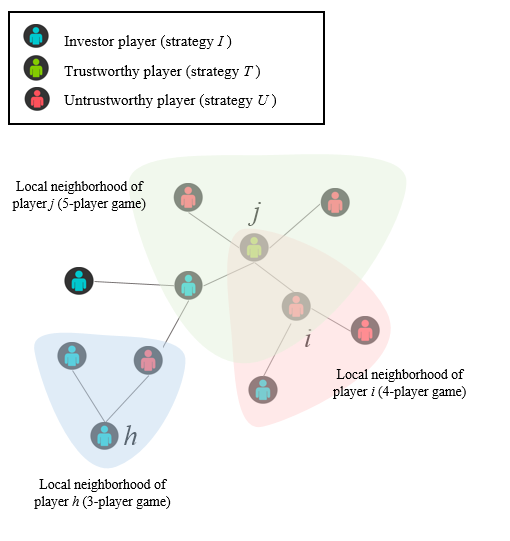}
\caption{An illustration of players' group diversity in the trust game.}
\label{fig:graph_local}
\end{figure}

Initially, the agents' strategies are assigned at random. Let us consider a local neighborhood with a focal agent, $i$, as shown in Figure \ref{fig:graph_local}. The game is played between $i$ and all its direct neighbors. In the neighborhood, there are $k_I$ investors, $k_T$ trustworthy trustees, and $k_U$ untrustworthy trustees. The equality $k_I + k_T + k_U = k_i$ must always be fulfilled for consistency's sake, where $k_i$ is the degree (number of connections) of focal node $i$. 

By the rules of the game, every investor releases a unit of payoff per time step. Since all the agents in the neighborhood are playing the game as a group, this means there are a total of $k_I$ units. Each of the $k_T+k_U$ trustees receives an equal division of this total quantity: $k_I /(k_T + k_U)$. A trustworthy trustee (strategy $T$) then returns the received fund multiplied by $R_T$ to the investors: $R_T \, k_I /(k_T + k_U)$ (i.e., $R_T \, k_I/k_{TU}$ to each of the $k_I$ investors, where $k_{TU} = k_T+k_U$). Each of the trustworthy trustees also receives the same amount of fund as the investors. That is, the investment produces in total $2\, R_T\, k_I/k_{TU}$. On the other hand, untrustworthy trustees (strategy $U$) return nothing but keep for themselves the received fund multiplied by $R_U$: $R_U \, k_I/k_{TU}$. 

The parameter $r_{UT} \in (0,1) $ acts as the temptation to defect ratio (to be untrustworthy), defined by $r_{UT} = (R_U-R_T/R_T)$, and it is used to control the difficulty level of the trust game. $r_{UT}$ values must fulfill the restriction $1 < R_T < R_U < 2 \,R_T$. Taking into account the number of players in the group and $r_{UT}$, we can determine the net wealth of individual agents, based on their payoffs, and according to the strategy adopted by themselves and their direct neighbors. The net wealth $w_i$ of a focal player $i$ is obtained as follows:
\begin{equation}
\label{eq:payoffs}
w_i=\left\{\begin{array}{ll}
	\frac{R_T \cdot k_T^*}{k_{TU}^*} - 1,  & i=1,...,k_I \\
    & \mbox{ (investor)},\\\\
	\frac{R_T \cdot k_I^*}{k_{TU}^*}, & i=(k_I+1),...,(k_I+k_T) \\
    & \mbox{ (trustworthy)},\\\\
	\frac{(1+r_{UT}) \cdot R_T \cdot k_I^*}{k_{TU}^*}, & i=(k_I+k_T+1),...,pop \\
    & \mbox{ (untrustworthy)},
\end{array}\right.
\end{equation}
where $k_{I}^*$, $k_{T}^*$, and $k_{TU}^*$ are the numbers of investors, trustworthy trustees, and both trustworthy and untrustworthy trustees in the local neighborhood of $i$, including the focal agent $i$ itself. The net wealth of focal agent $i$ is 0 when there is no trustee in the neighborhood (i.e., $w_i =0$ when $k_{TU}^*=0$). We are interested in the global net wealth of the population $W$, calculated as $W = \sum^{pop}_{i=1} w_i$. After playing a game and calculating their payoffs, the agents have an opportunity to update their strategies according to the payoffs received. This operation is carried out according to one of the evolutionary update rules, which will be described next. 

\begin{figure*}
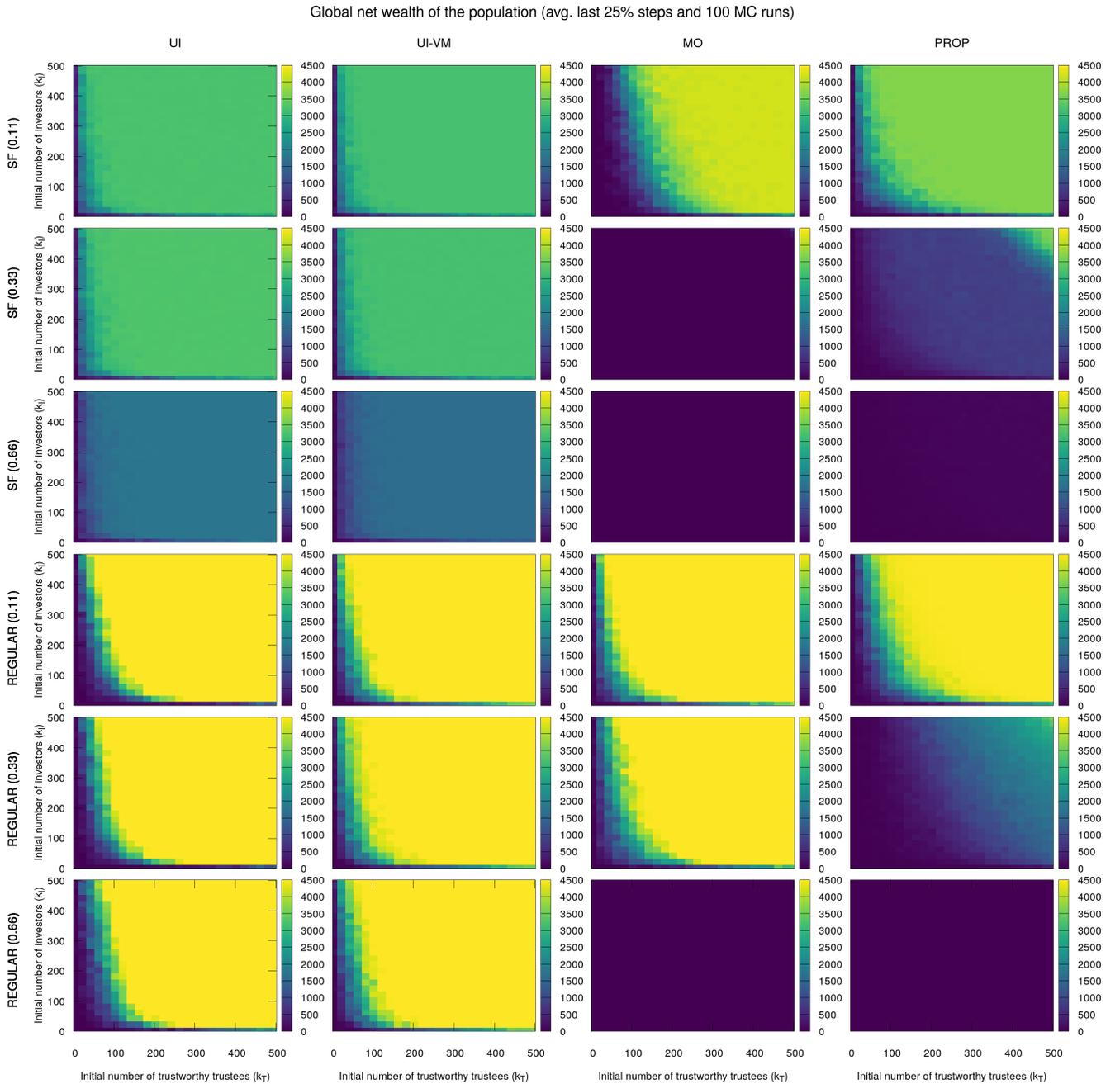

 \centering
\includegraphics[width=1\linewidth]{{{heatmaps_SA_SF_REGULAR_ALL}}}
\caption{Sensitivity analysis on different configurations of the initial population for the four considered update rules, two social network topologies, and three values of $r_{UT}$ that define the difficulty level of the game ($0.11, 0.33, 0.66$).}
\label{fig:panel_heatmaps_SA}
\end{figure*}

\subsection{\label{sec:update_rules} Update rules}

Agents decide which strategy to choose based on the population strategies in the previous time step (i.e., $t - 1$) and their payoff~\cite{Chica17JMR}. The strategy update follows an evolutionary procedure based on neighbor imitation that can be interpreted as information exchange within a social learning process~\cite{Nowak10}. To be more precise,  at time-step $t$, a focal agent $i$ (independently from its strategy) evaluates its previous payoff results in $t-1$ and decides whether to imitate a direct neighbor $j$'s strategy or not by using an evolutionary update rule. These rules of imitative nature are widely employed in the literature and represent a situation where bounded rationality or lack of information forces players to copy (imitate) others' strategies~\cite{Schlag98}. In this work, we have considered the following four update rules because of their different features (i.e., some are deterministic and others stochastic; some of them evaluate all the neighbors and can make mistake while others not):

\begin{itemize}

\item Unconditional imitation (UI)~\cite{Nowak92}. A player $i$ directly copies the strategy of the best performing neighbor $j$ only if the payoff $w^{t-1}_j$ is better than $w^{t-1}_i$. UI is a deterministic and strategic rule based only on payoff maximization when imitating others.

\item Hybridization of UI and a voter model (UI-VM). By using this mechanism, players imitate others by hybridizing two rules: UI, a totally deterministic and payoff based rule, and the voter model~\cite{Holley75}, a random decision-making mechanism based on just picking the strategy of one of the neighbors, $j$, at random. As done in~\cite{Vilone14}, our model uses a probability parameter $q$ for players to select between the two update rules: the voter model is chosen as the update rule by player $i$ with probability $q$ while UI is chosen by player $i$ with probability $1-q$. We set $q = 0.1$ for our experiments after some preliminary analysis across a range of other values.

\item The Moran (MO) process~\cite{Moran62}. With this update rule, focal agent $i$ evaluates all its neighbors' payoffs and assigns probabilities to them. This update rule is also stochastic but based on imitation dependent on the payoffs. At each time step there are possibilities to imitate each of the neighbors. Player $i$ thus can make mistakes by imitating a worse-performing neighbor in terms of payoffs. 

\item Proportional imitation (PROP)~\cite{Helbing92}. It is a pairwise and stochastic update rule where players cannot make mistakes during the imitation process (i.e., never imitate players with a lower payoff value). Agent $i$ first evaluates if the individual net wealth of neighbor $j$ in the previous time step $t-1$ ($w^{t-1}_j$) is higher than its own ($w^{t-1}_i$). If the net wealth of $j$ is higher, $i$ will adopt the strategy of $j$ with a probability depending on the difference between their payoffs:
\begin{equation}
prob^t_i{j} = \frac{\max\{0, w^{t-1}_j - w^{t-1}_i\}}{\phi},
\label{eq:prop_imitation}
\end{equation} 
where $\phi=max_{w}-min_{w}$ is the maximum payoff distance between two players to have $prob^t_i{j}$ properly normalized. The minimum possible net wealth for the game $min_{w}$ is $-1$ when the focal agent is an investor and its neighborhood is formed by all untrustworthy trustees. The maximum possible net wealth for the game $max_{w}$ occurs when the focal agent is untrustworthy and all its neighbors are investors. In this case, $max_{w}$ is equal to $(1+r_{UT})R_Tk_I^*$.

\end{itemize}

\section{\label{sec:results_promotion_trust}GLOBAL NET WEALTH} 

\begin{figure*}
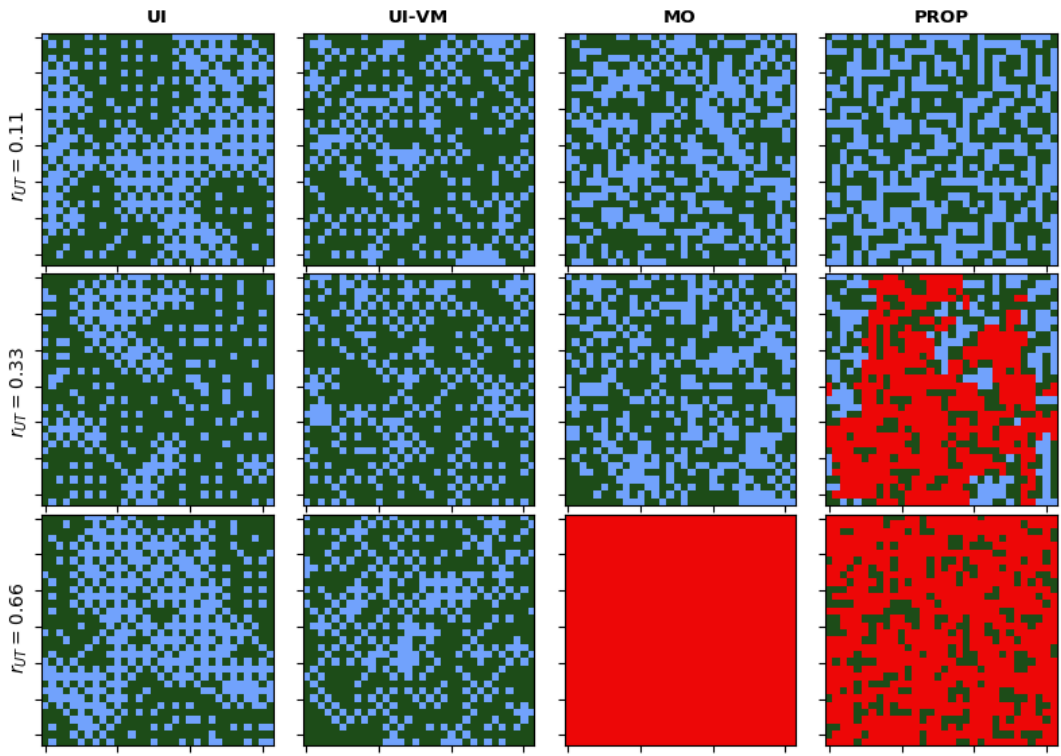

 \centering
\includegraphics[width=1\linewidth]{{{lattices_1024}}}
\caption{Snapshots of $32\times32$ lattices with 1,024 players for the four update rules and three different $r_{UT}$ values. Blue cells are players with strategy $I$, green cells are players with strategy $T$, and red cells are players with strategy $U$.}
\label{fig:lattice_1024}
\end{figure*}

\begin{figure}
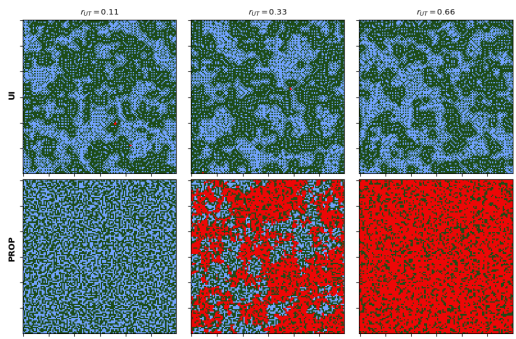

 \centering
\includegraphics[width=1\linewidth]{{{lattices_14400}}}
\caption{Snapshots of $120\times120$ lattices with 14,400 players for the UI and PROP update rules. Blue cells are players with strategy $I$, green cells are players with strategy $T$, and red cells are players with strategy $U$.}
\label{fig:lattice_14400}
\end{figure}

First, we will analyze the results of global net wealth of the population (i.e., the conditions and update rules to promote the highest sum of payoffs of all players of the population). The simulations were run for $5,000$ time steps until the model reached a stationary state using a population of 1,024 players. We set $R_T$ to 6, as in~\cite{Abbass16,Chica17IEEETEC}. The temptation to defect ratio $r_{UT}$ was set to three different values: $0.11, 0.33,$ and $0.66$, which cover three different trust scenarios: from an ``easier'' to a ``harder'' evolutionary trust game. All experiments were repeated for $100$ independent Monte Carlo (MC) realizations. 

We performed sensitivity analysis on the initial population conditions where the global net wealth was averaged over the last $25\%$ of the $5,000$ steps. Figure~\ref{fig:panel_heatmaps_SA} shows heatmaps of the results for the four update rules (individual columns of the panel) and three $r_{UT}$ values  with two social network topologies (see the six rows of the panel). Specifically, we show results for a regular lattice (REGULAR) and a SF network with average degree $\langle k \rangle=4$, generated using the Barabasi-Albert algorithm. The first three rows report on the net wealth using SF, with the game level rises from easy to a more difficult level ($r_{UT} = \{0.11, 0.33, 0.66\}$). The last three rows report on the same results when considering a regular lattice for the same difficulty game conditions.

\begin{figure*}
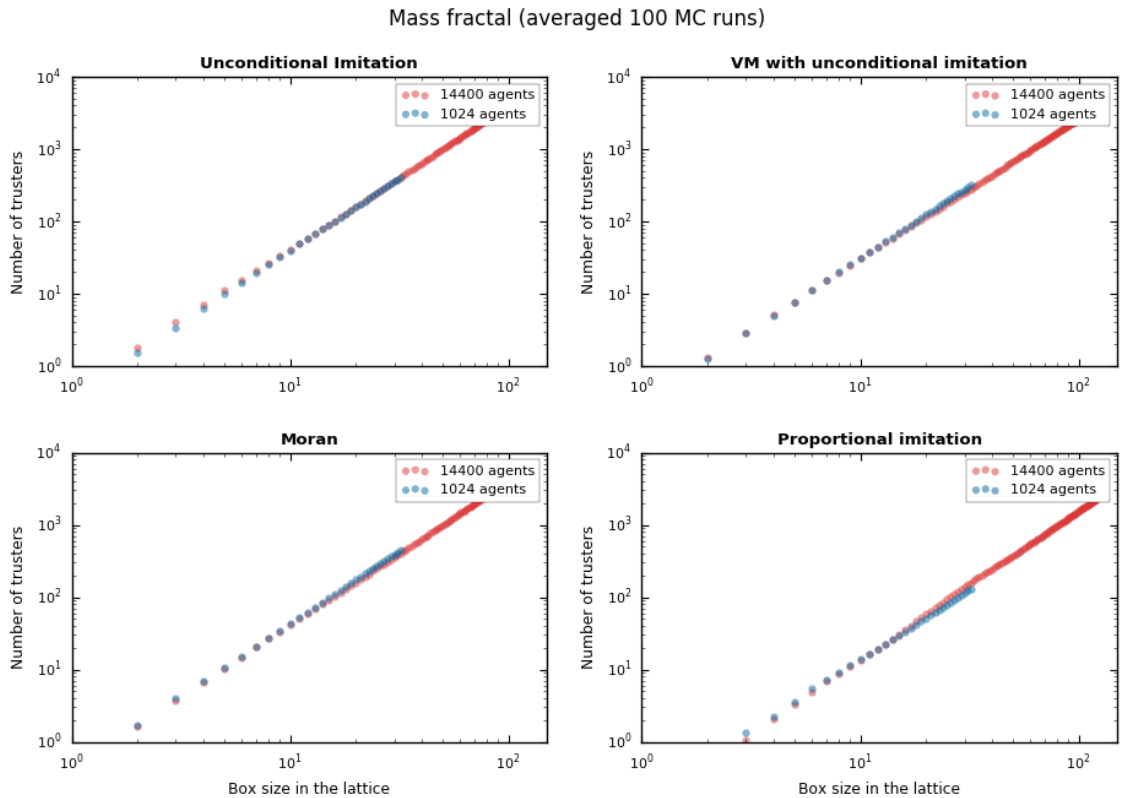

 \centering
\includegraphics[width=1\linewidth]{{{massFractals}}}
\caption{A mass study to show fractal structures for the four update rules in $32\times32$ (1,024 agents) and $120\times120$ (14,400 agents) lattices.}
\label{fig:mass_fractal}
\end{figure*}

We can see how, even for hard game conditions ($r_{UT} =0.66$), both UI and UI-VM can promote high ``cooperation'' levels under a wide range of initial population conditions when using a regular lattice (see the bottom left plots of the panel). Generally, both UI-based update rules (first two columns of the panel) perform equally in terms of promoting trust cooperation. However, when using a regular lattice, the use of UI-VM can generate a wider region of high global net wealth than when using the classical UI. 

Trust cooperation results obtained by UI-based rules, however, are not as good as MO and PROP update rules in those scenarios where we used a SF topology under easy game conditions (the first row of Figure~\ref{fig:panel_heatmaps_SA}). The MO rule is able to obtain higher global net wealth values but under narrower initial conditions (see the heatmap plot at position (1,3) of Figure~\ref{fig:panel_heatmaps_SA}). When the game becomes harder, both MO and PROP update rules cannot obtain high net wealth values when using a SF topology and trust cannot be promoted (same in the case of UI-based rules). See for example those heatmaps in the third row of the panel, which correspond to the case of $r_{UT}=0.66$. Both MO and PROP update rules are not able to promote non-zero global net wealth at the end of the simulations.


\section{\label{sec:results_analysis_dynamics}MODEL DYNAMICS}

In this section, we focus on the spatial and temporal dynamics of the game using different update rules. For all the analyses, we ran 100 independent MC realizations with an initial population formed by $30\%$ of players with strategy $I$, $25\%$ with strategy $T$, and $45\%$ with strategy $U$. We considered regular lattices of sizes $32\times32$ (i.e., 1,024 players) and $120\times120$ (i.e., 14,400 players) to better understand the spatial correlations between them.

\subsection{\label{sec:analysis_spatial}Spatial effects}

We start by showing spatial snapshots of the lattices after reaching an equilibrium state. Figure~\ref{fig:lattice_1024} shows, for each of the four update rules and three $r_{UT}$ values, the spatial layout of the lattices with the three possible strategies. Nodes in blue are those with strategy $I$, nodes in green are those with strategy $T$, and nodes in red are those with strategy $U$. We also show the spatial layout when considering the  $120\times120$ lattice (14,400 player) in Figure~\ref{fig:lattice_14400}.

\begin{figure*}
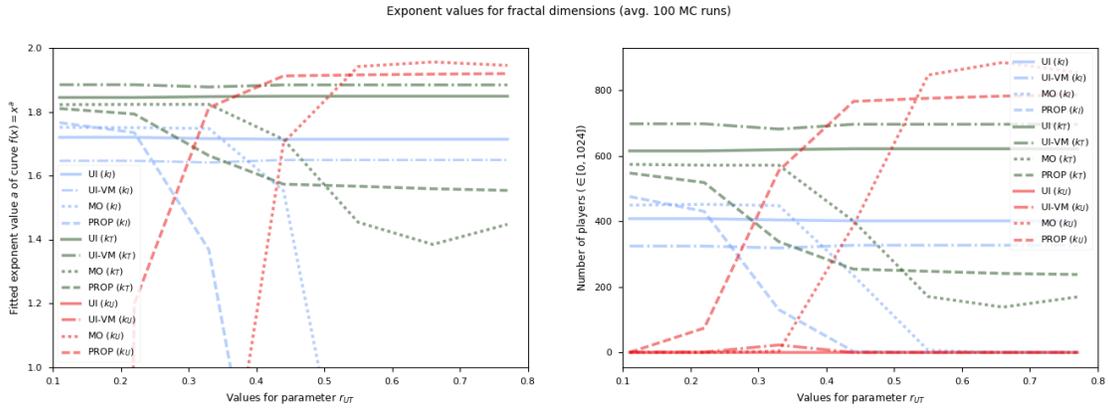

 \centering
\includegraphics[width=1\linewidth]{{{evolution_exps_strategies}}}
\caption{The right plot shows the number of players of each strategy ($k_I$, $k_T$, and $k_U$) at the end of the simulations. The left plot shows fitted exponents of the three fractal dimensions (i.e., the three playing strategies $k_I$, $k_T$, and $k_U$) for the four considered update rules. This experimentation was conducted using a $32\times32$ lattice.}
\label{fig:mass_fractal_additional}
\end{figure*}

For both lattice sizes we can see similar cluster and spatial formations. Also, the existing playing strategies are in line with the global net wealth observed in the previous section and Figure~\ref{fig:panel_heatmaps_SA}. UI and UI-VM rules, even with hard cooperation conditions, are able to eliminate all the $U$ players (only blue and green cells remain). With $r_{UT}=0.66$, however, untrustworthy players $k_U$ totally dominate the population when the MO rule is in place, and partially dominate the population when the PROP rule is used.

The spatial layout also suggests a fractal structure. This is especially clear when considering the UI rule, which provokes the emergence of ``organic'' clusters in the lattice. In order to further investigate the fractal properties of the game, we study, in a steady-state stage of the simulation, the relation of $k_I$ values (i.e., number of investors) when increasing a box size within the lattice. Figure~\ref{fig:mass_fractal} shows this relationship for both lattice sizes. Again, in this analysis we see that the number of players (i.e., the size of the lattice) does not affect the fractal properties of the game.

We have also fitted the exponents of these curves to find the fractal dimensions (i.e., the number of players playing each of the three strategies) and found their exponential values $a$ for a $f(x) = x^a$ function. These $a$ values are shown in Figure~\ref{fig:mass_fractal_additional} for $r_{UT}$ values ranging from $0.11$ to $0.77$ on the X axis. The right plot of the same figure shows the number of existing playing strategies to complement the fractal analysis. As we can observe in these plots, for all the update rules, the output of the spatial model presents a fractal structure with $a$ values between $1.6$ and $1.9$. This fractal structure is constant under different game difficulty conditions and for the three dimensions, although clearer for the number of investors $k_I$.

\begin{figure*}
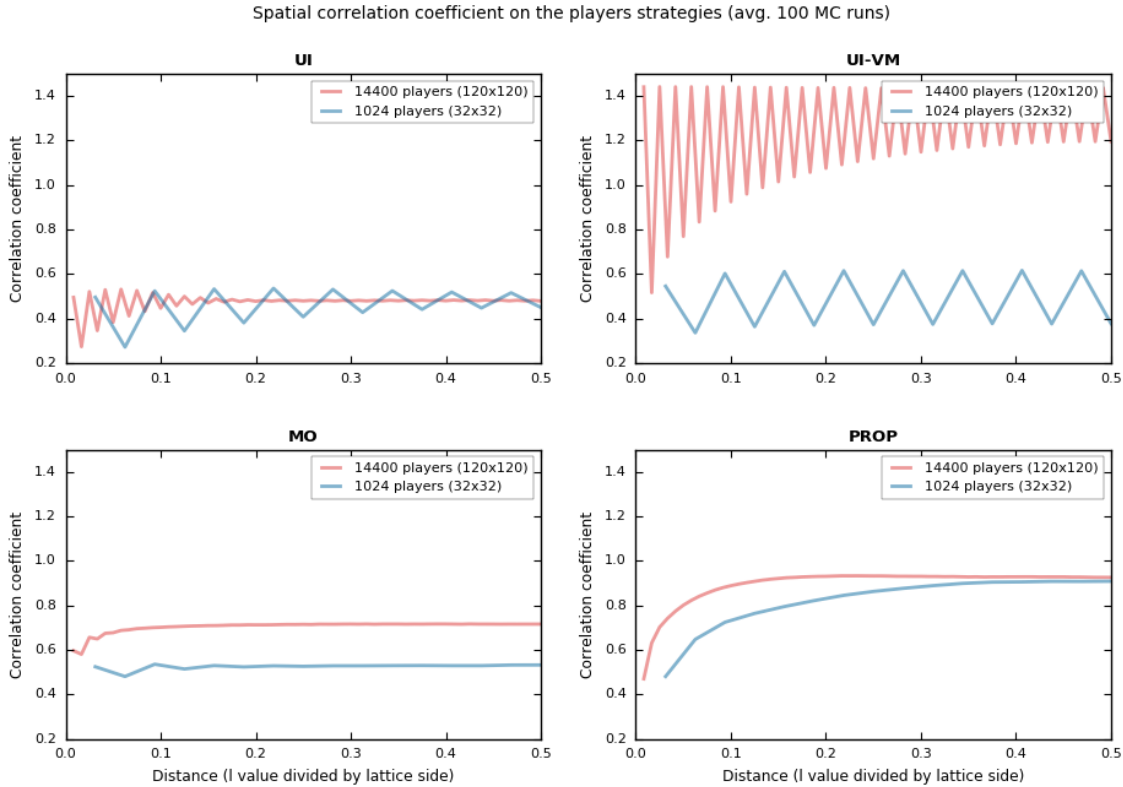

 \centering
\includegraphics[width=1\linewidth]{{{corrCoefs}}}
\caption{Spatial correlation coefficients between players in a lattice separated at several distances $l$ (both $32\times32$ and $120\times120$ regular lattices).}
\label{fig:corr_coefs}
\end{figure*}

The last spatial correlation analysis of this section consists of calculating the spatial correlation coefficients for all the players in the lattice located at a distance $l$. We follow a similar approach in~\cite{Fernandez14} by computing a spatial correlation function $G(l)$ as follows:
\begin{equation}
G(l) = \frac{1}{n} \sum_{\forall i,j / d(i,j)=l}(str(i) - str(j))^2,
\label{eq:g_l}
\end{equation} 
where $n$ is the number of pairs of agents at distance $l$ in the lattice, $d(i,j)$ is the distance in the lattice between players $i$ and $j$, and $str(i) \in \{1,2,3\}$ represents the three possible strategies of player $i$. Figure~\ref{fig:corr_coefs} shows all the $G(l)$ values for both $32 \times 32$ and $120 \times 120$ lattices for the four update rules. We find similar trends for both lattice dimensions (please note the X axis is relative to the lattice size), which again corroborates similar model behavior independent of the number of players. We observe cyclic oscillatory behavior when using one of the two UI-based rules (top plots of the figure). 

\subsection{\label{sec:analysis_temporal_dynamics}Temporal correlations}

\begin{figure*}
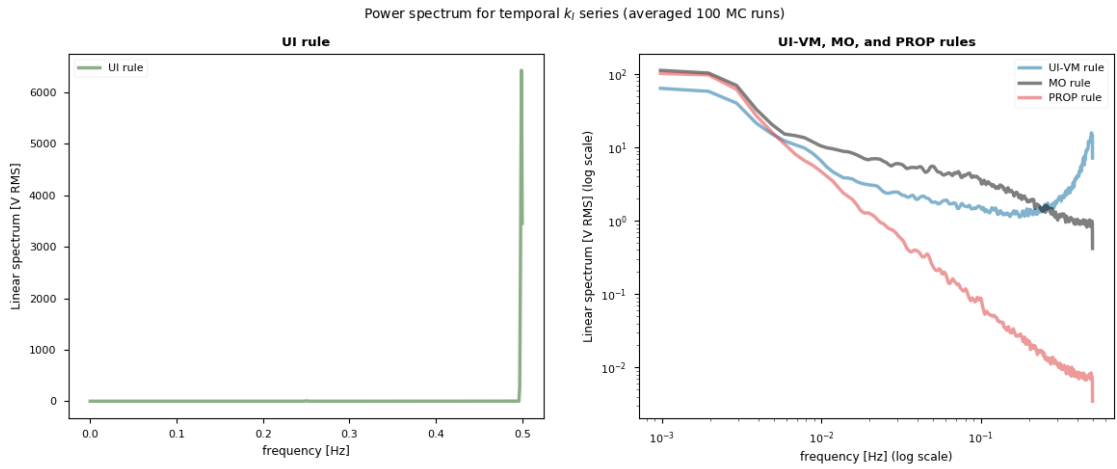

 \centering
\includegraphics[width=1\linewidth]{{{power_spectrum_rules_summarized}}}
\caption{Power spectrum of the time series evolution in $k_I$ for a $32 \times 32$ lattice based on the UI (left plot) and UI-VM, MO, and PROP rules (right plot).}
\label{fig:power_spectrum}
\end{figure*}

In this section, we explore temporal correlations of the model. Figure~\ref{fig:power_spectrum} shows, for each update rule, the power spectrum of the signal, which refers to the total number of players playing strategy $I$ (i.e., $k_I$) at each step of the simulation. This signal is obtained when the simulation reaches a steady state. The left-hand plot shows the power spectrum for the UI rule in a linear scale. We see that the signal has a high frequency of $0.5 \, Hz$, which means that there is a complete temporal correlation with a period of 2 steps. This is because of the oscillatory behavior of UI within the model, also shown in the spatial analysis in the next section.

More interestingly, the plot on the right shows the power spectrum of the other three update rules in a log-log scale. The three spectra of the rules have a power-law distribution. This distribution is especially clear for the PROP update rule, which shows a linear drop. These power-law distributions of the spectra of the rules show that the model's output has many different low frequencies. Therefore, when using non-deterministic update rules, the model has a long-term memory (i.e., long temporal correlations). This behavior is not common in this kind of models, as many of them exhibit stochastic dynamics, where there is only a short memory and the decision does not depend on previous history~\cite{Torney11}. Finally, we also see how the power spectrum of the UI-VM rule (blue line) has a high-frequency peak because of the deterministic nature of UI. 

\begin{figure*}
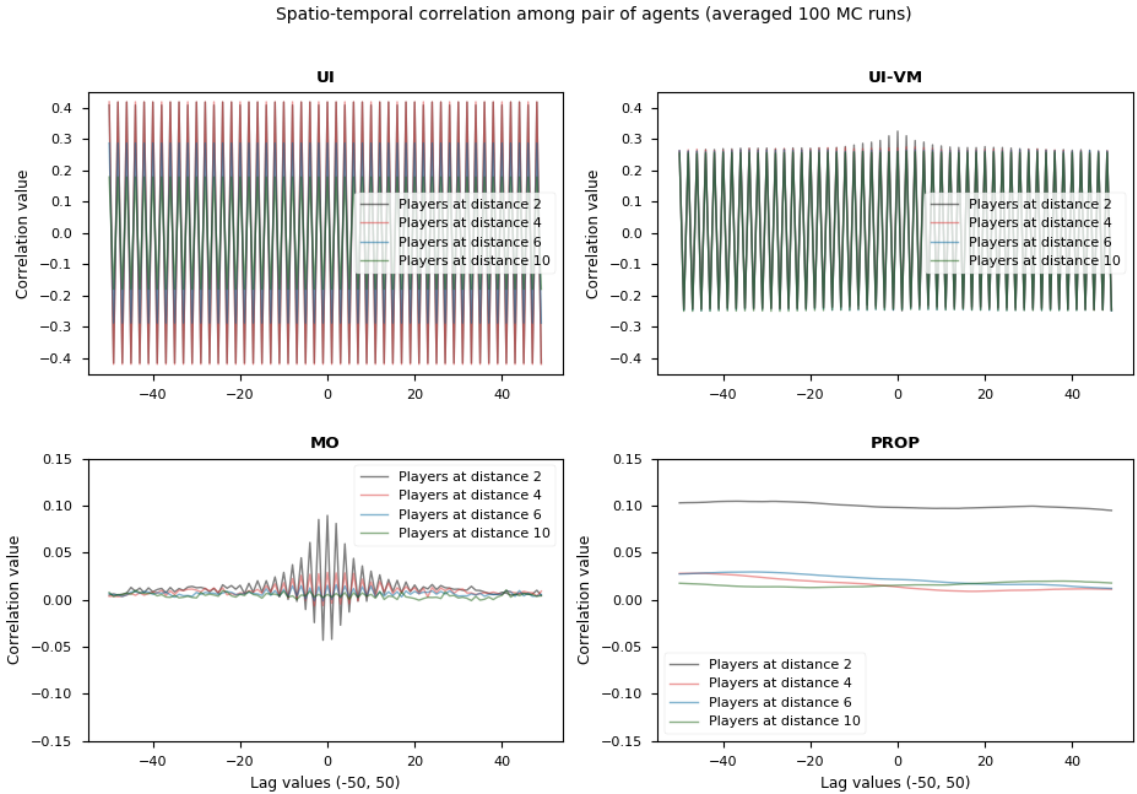

 \centering
\includegraphics[width=1\linewidth]{{{corr_agents_at_distance}}}
\caption{Spatio-temporal correlations between a focal player and players at distances 2, 4, 6, and 10 in a lattice.}
\label{fig:spatio_temporal_corr}
\end{figure*}

\subsection{\label{sec:analysis_spatial_dynamics}Spatio-temporal correlations}

We have calculated the spatio-temporal correlations among pairs of players in the lattice as the final stage of our experimentation analysis. Figure~\ref{fig:spatio_temporal_corr} shows plots with the time-lag Pearson correlation values of four players' time series with respect to a focal player. The calculation process was done as follows. We first set a focal agent at random and four additional players situated at a distance of 2, 4, 6, and 10 in the lattice with reference to the focal one. The time series strategy evolution of the focal agent and the other four players in the last 50\% of the steps are obtained. Later, the focal agent time series is compared to the other four players' time series to find spatio-temporal correlations. 

Here, we can see similar behavior to that observed when analyzing the spatial and temporal correlations independently in the previous sections. There are high correlation values for the UI deterministic rules (both UI and UI-VM) and these correlations are cyclic and for all the considered distances. This is related to the oscillatory behavior seen for the UI rule. In the case of the other two update rules (please note a lower Y scale for the MO and PROP plots), correlation values are lower. Proportional imitation (the bottom right plot of Figure~\ref{fig:spatio_temporal_corr}) has the highest correlation with the closest players (i.e., at distance 2). When using the MO update rule, correlations tend to become a null value when increasing the lag time between the series. As happened with the PROP rule, there is a big difference between the correlation values of closer players and those at 6 or 10 nodes, where there is not any spatio-temporal correlation.

\section{\label{sec:conclusions}DISCUSSION}

In this work, we have shown the dynamics of different update rules for a multi-player trust game of three strategies. First, we looked at which rules and social network topologies can obtain the highest trust cooperation among players in the population. Based on the simulation results, we can divide the update rules into two groups: (1) the UI strategy and its hybridized update rule with a voter model; and (2) the PROP and MO rules. PROP and especially the MO update rule obtain the best results in terms of global trust promotion when the game is easy ($r_{UT}=0.11$), and the network topology is SF. When the game is difficult, however, both UI-based rules are able to better promote trust and obtain higher global net wealth values for very different initial population conditions. In contrast, MO and PROP are not able to promote trust under difficult game conditions. With respect to the network topology, using a regular lattice enables better trust promotion leading to higher global net wealth.

Apart from evaluating trust promotion by the rules, we studied the spatial and temporal model dynamics. We found that the UI-based rules form spatial clusters of sponge-like areas. These patterns are created to enable the maximization of groups of trustworthy trustees (strategy $T$) with at least one investor (strategy $I$). More interestingly, we found that not only UI but all the update rules generate a fractal space on the lattice when reaching a steady-state stage. Generally, spatial correlations are important and occur mainly in closer players within the lattice. By studying the power spectrum of the model's output we also observed that when introducing stochasticity in the rules, a long-term memory appears in the system. This means that for the UI-VM, MO, and PROP rules, there are signals at low frequencies and their spectra are of power-law nature. 

Another interesting behavior is that the system is not frozen when reaching a steady state for deterministic rules such as the UI rule. There is a phenomenon related to ``switching'' or cyclic oscillatory behavior (e.g., from investor strategy $I$ to trustworthy trustee $T$ or the other way round). From the temporal and spatio-temporal analyses we found these oscillations and maximum correlations with a 2-step period (a high frequency of $0.5Hz$). It is also clear that all of these switching nodes must have more than one connection to create a ``chain'' of switching nodes. The latter switching behavior was not observed in the PROP update rule. We think the reason for not having this behavior when using the PROP rule is because it is the only rule that only evaluates one neighbor to make a decision. When considering rules such as UI, UI-VM, and MO, if one neighbor $j$ changes the strategy, the decision of focal player $i$ changes totally, as all the neighbors are evaluated to make a decision to imitate. 

These important findings about the hidden dynamics and spatio-temporal correlations in the game show the complexity of managing trust in social networks in the long run. They also open doors for future research on evolutionary game theory in general and evolutionary trust games in particular.



%

\end{document}